\RequirePackage{ifpdf}
\ifpdf 
\documentclass[pdftex]{sigma}
\else
\documentclass{sigma}
\fi

\begin{document}

\allowdisplaybreaks

\renewcommand{\thefootnote}{$\star$}

\renewcommand{\PaperNumber}{013}

\FirstPageHeading

\ShortArticleName{Shifted Riccati Procedure: Application to Conformal Barotropic FRW Cosmologies}

\ArticleName{Shifted Riccati Procedure: Application to Conformal\\ Barotropic FRW Cosmologies\footnote{This
paper is a contribution to the Proceedings of the Workshop ``Supersymmetric Quantum Mechanics and Spectral Design'' (July 18--30, 2010, Benasque, Spain). The full collection
is available at
\href{http://www.emis.de/journals/SIGMA/SUSYQM2010.html}{http://www.emis.de/journals/SIGMA/SUSYQM2010.html}}}

\Author{Haret C. ROSU~$^\dag$ and Kira V. KHMELNYTSKAYA~$^\ddag$}

\AuthorNameForHeading{H.C. Rosu and K.V. Khmelnytskaya}

\Address{$^\dag$~IPICyT, Instituto Potosino de Investigacion Cientifica y Tecnologica, \\
\hphantom{$^\dag$}~Apdo Postal 3-74  Tangamanga, 78231 San Luis Potos\'{\i}, Mexico}
\EmailD{\href{mailto:hcr@ipicyt.edu.mx}{hcr@ipicyt.edu.mx}}

\Address{$^\ddag$~Universidad Aut\'onoma de Quer\'etaro, Centro Universitario,\\
\hphantom{$^\ddag$}~Cerro de las Campanas s/n
C.P. 76010 Santiago de Quer\'etaro, Qro. Mexico}
\EmailD{\href{mailto:khmel@uaq.mx}{khmel@uaq.mx}}

\ArticleDates{Received November 30, 2010, in f\/inal form January 28, 2011; Published online February 02, 2011}

\Abstract{In the case of barotropic FRW cosmologies, the Hubble parameter in conformal time is the solution of a simple Riccati equation of constant coef\/f\/icients. We consider these cosmologies in the framework of nonrelativistic supersymmetry that has been ef\/fective in the
area of supersymmetric quantum mechanics. Recalling that 
Faraoni~[\emph{Amer. J. Phys.} \textbf{67} (1999), 732--734]
showed how to reduce the barotropic FRW system of dif\/ferential equations to simple harmonic oscillator dif\/ferential equations, we set the latter equations in the supersymmetric approach and divide their solutions into two classes of `bosonic' (nonsingular) and `fermionic' (singular) cosmological zero-mode solutions. The fermionic equations can be considered as representing cosmologies of Stephani type,  
i.e., inhomogeneous and curvature-changing in the conformal time.
We next apply the so-called shifted Riccati procedure by introducing a constant additive parameter, denoted by $S$, in the common Riccati solution of these supersymmetric partner cosmologies. This leads to barotropic Stephani cosmologies with periodic singularities in their spatial curvature indices
that we call $\mathcal{U}$ and $\mathcal{V}$ cosmologies, the f\/irst being of bosonic type and the latter of fermionic type.
We solve completely these cyclic singular cosmologies at the level of their zero modes showing that an acceptable shift parameter should be purely imaginary, which in turn introduces a parity-time (PT) property of the partner curvature indices.}

\Keywords{factorization; shifted Riccati procedure; barotropic FRW cosmologies; cosmological zero-modes}

\Classification{81Q60}

\renewcommand{\thefootnote}{\arabic{footnote}}
\setcounter{footnote}{0}

\section{Introduction}

Riccati nonlinear equations have many applications in physics 
and their solutions under the name of superpotentials play an important role in supersymmetric quantum mechanics \cite{susyqm} (for a chronological review see~\cite{Rosu-quant-ph/9809056}).
In the cosmological framework, the supersymmetric methods have been recently reviewed in the book of Moniz \cite{bookM} and in general the occurrence of Riccati equations is well documented, see for example~\cite{rn}.
With the cosmological application that we will consider in this paper in mind, we write the Riccati equation in the form
\begin{gather}  \label{ricco1}
R^{\prime}+ cR^2+f=0,
\end{gather}
where $c$ is a real constant and $f$ is a function of the independent variable. It is well known that particular solutions of the Riccati equation enter the factorization brackets of second order linear dif\/ferential equations (usually with $c=1$)
\begin{gather}
\left( \frac{d}{dt}+cR\right) \left( \frac{d}{dt}-cR\right)u=0 \quad  \Leftrightarrow \quad   u''-c(cR^2+R')u=0 \quad   \Leftrightarrow \quad   u''+cfu=0 .
\label{fact1}
\end{gather}%
The connections $R=\frac{1}{c}\frac{u'}{u}$ or $u=e^{c\int^t R}$ between the particular solutions of the two equations are also basic results of the factorization method together with the construction of the so-called supersymmetric partner equation of equation (\ref{fact1}) obtained by reverting the order of the factorization brackets:
\begin{gather*}
\left( \frac{d}{dt}-cR\right) \left( \frac{d}{dt }+cR\right)v=0 \quad   \Leftrightarrow \quad    v''-c(cR^2-R')v=0 \quad   \Leftrightarrow \quad    v''+c(f+2R')v=0 .
\end{gather*}

\section{Shifted Riccati procedure}\label{sec2}

In previous works, we have introduced a supersymmetric procedure in which we examined the consequences of a constant shift of the Riccati solution \cite{k1, ro-06}, i.e.,
\[
R_S(t)=R(t)+S.
\]
$R_S$ obeys a Riccati equation of the form:
\[
R^{\prime}_{S}- 2cS R_{S}+cR^{2}_{S}+\big(f+cS^2\big)=0,
\]
which for $S=0$ turns into the simple form (\ref{ricco1}).
The corresponding linear second-order dif\/ferential equation obtained by substituting $R_S=\frac{1}{c}\frac{u_S'}{u_S}$ is:
\[
u_S''-2cSu_S'+\big(cf+c^2S^2\big)u=0
\]
and one can immediately see that a particular solution is $u_S=e^{cSt}u$. Similarly, $v_S= e^{cSt}v$ is a~particular solution of the supersymmetric equation
\[
v_S''-2cSv_S'+c\big(f+2R'+cS^2\big)v=0
\]
related to the non-standard Riccati equation
\[
-R^{\prime}_{S}- 2cS R_{S}+cR^{2}_{S}+\big(f+2R'+cS^2\big)=0.
\]

However, if we consider the shifted Riccati solution $R_S$ directly in the factorization brackets we obtain the following pair of supersymmetric equations:
\begin{gather}
\left( \frac{d}{dt}+cR_S\right) \left(\frac{d}{dt}-cR_S\right)U=0 \quad   \Leftrightarrow \quad U''+c(f-2cSR-cS^2)U  =0, \label{array1}\\
\left( \frac{d}{dt}-cR_S\right) \left(\frac{d}{dt}+cR_S\right)V=0 \quad   \Leftrightarrow \quad V''+c(f-2cSR-cS^2+2R')V  =0 .\label{array2}
\end{gather}
These equations reduce to the equations for $u$ and $v$ when $S=0$. Nevertheless, their solutions are not connected to the solutions $u$ and $v$ in the same simple way that $u_S$ and $v_S$ are. In the rest of the paper, we will present in full detail the solutions of the latter equations in the interesting cosmological case of barotropic FRW cosmologies.

\section{Cosmological Riccati equation in FRW barotropic cosmology}\label{sec3}

 In 1999, Faraoni \cite{Far} showed that the well-known comoving
Einstein--Friedmann dynamical equations of barotropic FRW cosmologies can be
turned into a single Riccati equation for the Hubble parameter in conformal time ${\cal H}(\eta)$ of the form
\begin{gather}\label{ricc}
{\cal H}^{\prime}+\tilde{\gamma}{\cal H}^{2}+\kappa\tilde{\gamma} =0,
\end{gather}
where henceforth the prime and also $\frac{d}{d \eta}$ stand for the
derivative with respect to $\eta$, $\tilde{\gamma}=\frac{3}{2}\gamma-1$ is related to the adiabatic index $\gamma$ of the cosmological f\/luid
and $\kappa=\pm 1$ is the curvature parameter for the closed and open universe, respectively. Notice that for $\gamma=\frac{2}{3}$ equation (\ref{ricc}) reduces to ${\cal H}^{^{\prime}}=0$ which is not a Riccati equation and therefore the arguments of this paper do not apply in this case. Moreover, equation (\ref{ricc}) is just the Riccati equation of the harmonic oscillator:
\begin{gather*}
\dot{R}+\omega_{0}R^{2}+\omega_{0}=0,
\end{gather*}
if one sets $\tilde{\gamma}\equiv\omega_{0}$ for the closed universe case.
For the open case the analogy is with the up-side down harmonic oscillator.
The main dif\/ference resides merely in the independent variable, which in the
cosmological framework is the conformal time whereas in the harmonic
oscillator case is the usual Newtonian time.
In the following, we will consider only the $\kappa=1$ case since it allows us to focus on the periodic features of the problem.

Usually the conformal Hubble parameter is def\/ined as the logarithmic derivative of the
scale factor of the universe, ${\cal H}(\eta)=\frac{a^{\prime}}{a}$, which is related to the observational (comoving) Hubble parameter through
${\cal H}(\eta)=a(t)H(a(t))$, for a discussion of this issue see \cite[p.~96]{Paranjape09}. Here, we def\/ine the conformal Hubble parameter directly in terms of the zero modes, i.e., the particular solution~$u$, as
${\cal H}_u(\eta)=\frac{1}{\tilde{\gamma}}\frac{u^{\prime}}{u}$. Substituting ${\cal H}_u$ in
equation (\ref{ricc}) leads to the very simple (harmonic oscillator) second order dif\/ferential equation
\begin{gather}
\label{w}
u^{\prime\prime}+\kappa_{u}\tilde{\gamma}^{2}u=0, \qquad\kappa_{u}=1.
\end{gather}

Employing a terminology used in supersymmetric quantum mechanics we have previously called the $u$ modes as bosonic zero modes \cite{ro-06}.
They are $\tilde{\gamma}$ powers of the scale factor parameters~$a(\eta)$, i.e.,
$u=a^{\tilde{\gamma}}(\eta)$. Faraoni used the following particular
solutions:
\[
u_{1} \sim\cos\tilde{\gamma}\eta\qquad\Rightarrow\qquad a_{1u}(\eta) \sim
u_{1}^{1/\tilde{\gamma}} .
\]
Using $u_{1}$ in the def\/inition of ${\cal H}_u$ one gets ${\cal H}_{u_{1}}=-
\tan\tilde{\gamma}\eta$.


 A class of cosmologies with inverse scale factors with respect to the standard barotropic ones
but with a specif\/ic conformal time dependence of the curvature index 
can be obtained by considering the supersymmetric partner
equation of equation (\ref{w}), which is given by \cite{ro-06}
\begin{gather*}
v^{\prime\prime} +\kappa_{v}(\eta)\tilde{\gamma}^{2}v=0,
\end{gather*}
where
\begin{gather}\label{geta}
\kappa_{v}(\eta) =-\big(1+2\tan^{2} \tilde{\gamma}\eta\big)
\end{gather}
denotes the conformal time dependent supersymmetric partner curvature index of
fermionic type associated through the mathematical scheme to the constant bosonic curvature index.

A particular fermionic solution $v$ is of the following type
\[
v_{1} =\frac{\tilde{\gamma}}{\cos\tilde{\gamma}\eta} \qquad\Rightarrow\qquad
a_{1v}(\eta) \sim v_{1}^{1/\tilde{\gamma}}.
\]

We can see that the $u$ and $v$ barotropic cosmologies are dual to each other from the standpoint of these particular solutions, in the sense that
$u_1v_1=\tilde{\gamma}$ and therefore the geometric mean of their scale parameters
\[
a_g = \left(a_{1u}a_{1v}\right)^{1/2}=(\tilde{\gamma})^{1/2\tilde{\gamma}}
\]
is constant. Thus, a joint evolution of a $u$ cosmology of constant curvature index and a $v$
cosmology of the time-dependent curvature index (\ref{geta}) is stationary in conformal time from the standpoint of their geometric mean scale parameter $a_g$. Since the fermionic cosmology is of time-dependent (periodic) curvature index, it can be also called a Stephani type cosmology~\cite{steph} (for a review see~\cite{Lorenz-Petzold}), although in conformal time. The fermionic metric is of the form:
\[
ds^2=a_{1v}^2(\eta)\left[ -d\eta^2+\frac{dr^2}{1-\kappa_{v}(\eta)r^2}+r^2d\Omega^2\right].
\]
This metric should be thought of as an averaged metric in an inhomogeneous cosmology and so it does not even have to satisfy the Einstein f\/ield equations \cite{Flanagan08}, although we prefer to keep this constraint for philosophical reasons (that is why we consider to be of Stephany type). The only thing it has in common with the standard FRW model is the Riccati solution. Such metrics, with other time-dependent curvature indices have been used to mimic the backreaction of small scale density perturbations on the large scale spacetime geometry~\cite{Flanagan08, Larena09}.

\section{Barotropic conformal-time Stephani-type cosmologies\\ from a constant shift of the Hubble parameter}\label{sec4}

We would like now to investigate the implications of the constant shifting $S$ of the conformal Hubble parameter, according to the procedure discussed in Section~\ref{sec2}. This means that we will build Schr\"odinger equations based on the Riccati solution
${\cal H}_{S}(\eta)={\cal H}_{u_{1}}(\eta)+S$, similar to the equations (\ref{array1}) and (\ref{array2})
\begin{gather}
\left(  \frac{d}{d\eta}+\tilde{\gamma}{\cal H}_{S}\right)  \left(  \frac{d}{d\eta}-\tilde{\gamma}{\cal H}_{S}\right)
\mathcal{U}=0\quad \Leftrightarrow\nonumber\\
\mathcal{U}^{\prime\prime}+\mathcal{K}_{u}(\eta){\tilde{\gamma}^{2}}
\mathcal{U}=0 ,
\qquad  {\cal K}_{u}(\eta)=1-S^{2}+2S
\tan\tilde{\gamma}\eta
\label{calU}
\end{gather}
and
\begin{gather}
\left(  \frac{d}{d\eta}-\tilde{\gamma}{\cal H}_{S}\right)  \left(  \frac{d}{d\eta}+\tilde{\gamma}{\cal H}_{S}\right)
\mathcal{V}=0
\quad \Leftrightarrow\nonumber\\
\mathcal{V}^{\prime\prime}+\mathcal{K}_{\nu}(\eta){\tilde{\gamma}^{2}}\mathcal{V}=0 ,\qquad {\cal K}_{\nu}(\eta)=-
1-S^{2}+2S
\tan\tilde{\gamma}\eta-2\tan^{2}\tilde{\gamma}\eta
.\label{calV}
\end{gather}

Equations (\ref{calU}) and (\ref{calV}) def\/ine two new classes of barotropic-like cosmological universes that we call $\mathcal{U}$ and
$\mathcal{V}$ cosmologies, respectively. One can also write averaged conformal-like metrics with the corresponding variable curvature indices. These two modif\/ied cosmologies are periodic, of period $T=\frac{\pi}{\tilde{\gamma}}$, and have the same conformal Hubble parameter given by:
\begin{gather*}
{\cal H}_{\mathcal{U}}(\eta)={\cal H}_{\mathcal{V}}(\eta)={\cal H}_{S}(\eta)=-\frac{\sin\tilde{\gamma
}\eta}{\cos\tilde{\gamma}\eta}+S.
\end{gather*}
Obviously, ${\cal H}_{\mathcal{U},\mathcal{V}}\rightarrow {\cal H}_{u_{1}}$, when $S\rightarrow0$. Both cosmologies have periodic singularities in their time-dependent curvature indices. In addition, ${\cal H}_S(0)=S$, i.e., $S$ serves as an initial condition for the conformal Hubble parameter of these cosmologies.

The linear independent solutions $\mathcal{U}_{1}$ and $\mathcal{U}_{2}$ have the following form
\begin{gather}
\mathcal{U}_{1}(\eta)=e^{-i\Omega_{S}\eta}\,{}_{2}F_{1}\left(  1,-iS;2-iS;-e^{-2i\tilde{\gamma}\eta}\right)  ,\label{2-fcfin1}
\end{gather}
or $\mathcal{U}_{1}(\eta)=e^{-S\tilde{\gamma}\eta}\left[z \,{}_{2}F_{1}\left(  1,-iS;2-iS;-z^2\right)\right]$, where $z=e^{-i\tilde{\gamma}\eta}$, $\Omega_{S}=\tilde{\gamma}(1-iS)$ and $_{2}F_{1}\left(  a,b;c;z\right) $ is the hypergeometric function, and
\begin{gather}
\mathcal{U}_{2}(\eta)= e^{i\Omega_{S}\eta}\big(2\cos^{2}\tilde{\gamma} \eta-i\sin2\tilde{\gamma} \eta\big)\sim e^{S\tilde{\gamma}\eta}\cos \tilde{\gamma}\eta .\label{2-fcfin2}
\end{gather}
We also notice that the simple convergence condition $\Re(a+b-c)<0$ for the hypergeometric series is fulf\/illed for all real values of $\eta$.

On the other hand, the linear independent solutions $\mathcal{V}_1$ and $\mathcal{V}_2$ are given by:
\begin{gather}
\mathcal{V}_{1}(\eta)   =\frac{e^{-i\Omega_{S}\eta}}{(2\cos^{2}\tilde{\gamma}\eta-i \sin2\tilde{\gamma} \eta)}  ,\label{g1} 
\end{gather}
or $\mathcal{V}_{1}(\eta)=\frac{e^{-S\tilde{\gamma}\eta}}{2\cos \tilde{\gamma}\eta}$ and
\begin{gather}
\mathcal{V}_{2}(\eta)    = \tilde{\gamma}^{2}e^{i\Omega_{S}\eta}\left[ 
(1-i\tan\tilde{\gamma}\eta )+iS(1-iS)\big(2\cos^{2}\tilde{\gamma} \eta-i \sin2\tilde{\gamma} \eta\big) -2iS
\right]  ,\label{g2}%
\end{gather}
or $\mathcal{V}_{2}(\eta)=2\tilde{\gamma}^2e^{S\tilde{\gamma}\eta}\left[\frac{1}{2\cos \tilde{\gamma}\eta}+S(S\cos \tilde{\gamma}\eta +\sin \tilde{\gamma}\eta)\right]$.

\begin{figure}[t]
\centering
\includegraphics{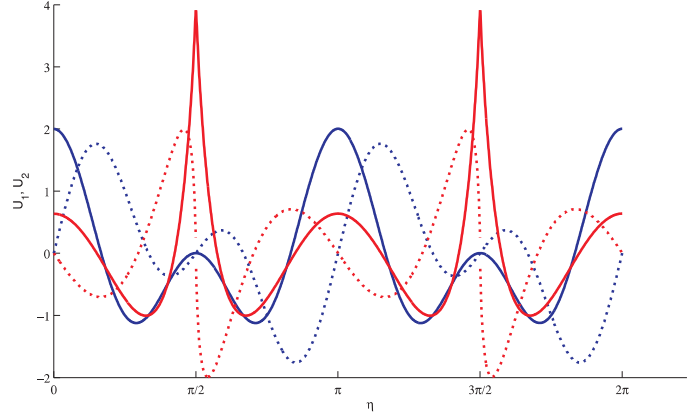}
\caption{The real (solid lines) and imaginary (dotted lines) parts of the
periodic zero modes $\mathcal{U}_{1}(\eta)$ (in red) and $\mathcal{U}_{2}
(\eta)$ (in blue) for the shift parameter $S=3i$.}\label{figureS1}
\end{figure}

\begin{figure}[t]
\centering
\includegraphics{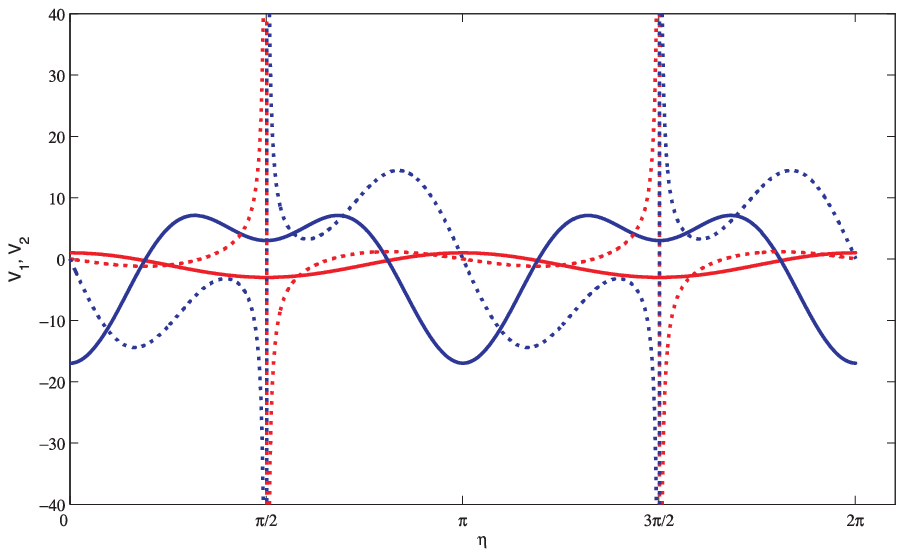}
\caption{The real (solid line) and imaginary (dotted lines) parts of the
periodic zero modes $\mathcal{V}_{1}(\eta)$ (in red) and $\mathcal{V}_{2}
(\eta)$ (in blue) for the shift parameter $S=3i$.}\label{figureS2}
\end{figure}

\begin{figure}[th!]
\centering
\includegraphics{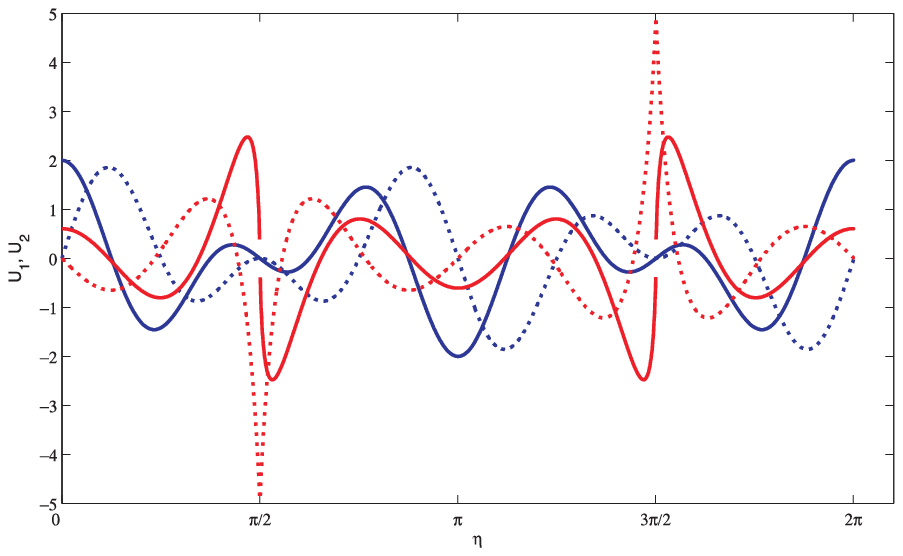}
\caption{The real (solid lines) and imaginary (dotted lines) parts of the
antiperiodic zero modes $\mathcal{U}_{1}(\eta)$ (in red) and $\mathcal{U}
_{2}(\eta)$ (in blue) for the shift parameter $S=4i$.}
\label{figureS3}
\end{figure}

\begin{figure}[t]
\centering
\includegraphics{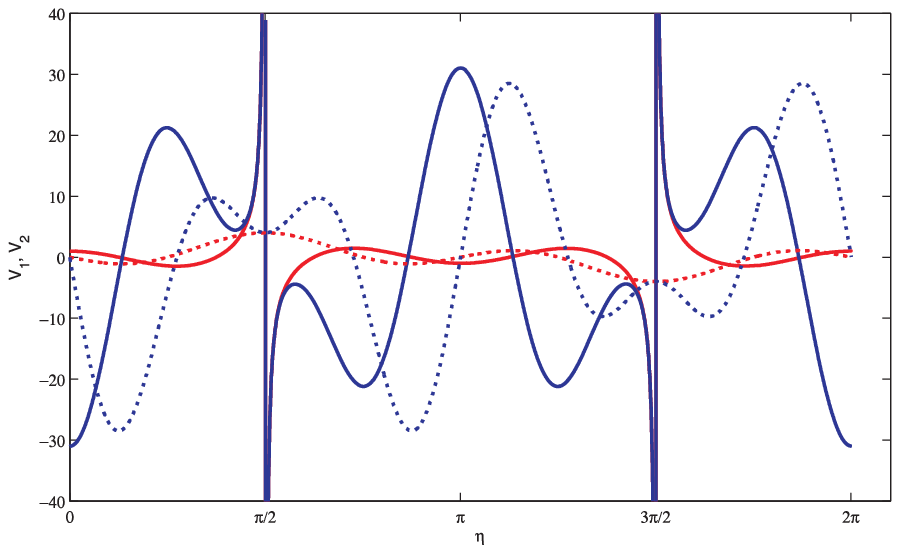}
\caption{The real (solid lines) and imaginary (dotted lines) parts of the
antiperiodic zero modes $\mathcal{V}_{1}(\eta)$ (in red) and $\mathcal{V}
_{2}(\eta)$ (in blue) for the shift parameter $S=4i$.}\label{figureS4}
\end{figure}

 One can notice immediately that we have written the solutions (\ref{2-fcfin1}), (\ref{2-fcfin2}) and (\ref{g1}), (\ref{g2}) in a~convenient Floquet--Bloch form that allows us to make the following comments.
The parameter~$S$ af\/fects only the period of the phases $e^{\pm i\Omega_{S}\eta}$ of the solutions but not that of their periodic part. The solutions (\ref{2-fcfin1}), (\ref{2-fcfin2}) and the Bloch factors  $e^{\pm i\Omega_{S}\eta}$ in (\ref{g1}), (\ref{g2}) are bounded if and only if the
``quasifrequency'' $\Omega_{S}$ has a real value, or equivalently
\begin{gather*}
\widetilde{\gamma}(1-iS)\in\Re .
\end{gather*}
Taking into account that $\widetilde{\gamma}\in\Re$, the last condition implies that $S$ is pure imaginary, i.e., $S=is$, $s\in\Re$. Thus, equation(\ref{calU}) has bounded solutions $\forall\, s\in\Re$.
Notice that for a purely imaginary shift parameter, the curvature indices $\mathcal{K}_{u}$ and $\mathcal{K}_{\nu}$ are related through $\mathcal{K}_{u,\nu}(-\eta)=\mathcal{K}^{*} _{u,\nu}(\eta)$, where $*$ denotes the complex conjugation operation. This means that we can have in this cosmological context the parity-conformal time (PT) symmetry.
Additionally, by inspecting the solutions
(\ref{2-fcfin1}), (\ref{2-fcfin2}) and (\ref{g1}), (\ref{g2}) we note that they
are periodic for $s_p=(2m-1)$ , $m=0,1,2,\dots$ and antiperiodic
for $s_a=2m$ , $m=\pm1,\pm2,\dots$. Plots of the solutions in the periodic case $m=1$ are displayed in Figs.~\ref{figureS1} and \ref{figureS2}, whereas the antiperiodic case $m=2$ can be found in Figs.~\ref{figureS3} and \ref{figureS4}. In general, the $\mathcal{U}$ modes are regular indicating that these shifted cosmologies are not sensitive to the singularities of their curvature
indices at the level of their zero modes.
On the other hand, the $\mathcal{V}$ cosmologies have periodic singularities in their imaginary parts but not in their real parts. In addition, the duality property is maintained for the pair of zero-modes $\mathcal{U}_{2}\mathcal{V}_{1} = {\rm const}$.

  The periodic case for $m=0$ gives the solutions
  \begin{gather*}
  \mathcal{U}_{1}^{0}(\eta)=2ie^{-i\tilde{\gamma}\eta}\sin\tilde{\gamma}\eta, \qquad
  \mathcal{U}_{2}^{0}(\eta)=2e^{-i\tilde{\gamma}\eta}\cos\tilde{\gamma}\eta
  \end{gather*}
  and
  \begin{gather*}
  \mathcal{V}_{1}^{0}
(\eta)=\frac{e^{i\tilde{\gamma}\eta}}{2\cos\tilde{\gamma}\eta} ,
\qquad \mathcal{V}_{2}^{0}(\eta)=\frac{\tilde{\gamma}^{2}e^{-i\tilde{\gamma}\eta}}{\cos\tilde{\gamma}\eta}-2\tilde{\gamma}^{2}.
\end{gather*}
The solution $\mathcal{U}_{1}(\eta)$ is further discussed in the Appendix.

Regarding the dualities we introduced here, it is worth emphasizing certain similarities with the superstring dualities~\cite{GV03} and the phantom duality~\cite{D03}. In the f\/irst case, there is an invariance of the action with respect to an inversion of the cosmological scale factor and special shifts of the value of the dilaton f\/ield ($a\rightarrow a^{-1}$ and $\Phi \rightarrow \Phi-6\ln a$, respectively). We also have the scale factor duality and we also considered a shift but of the conformal Hubble parameter. However, our context and mathematical techniques appear to be dif\/ferent. Moreover, the string action with such properties corresponds to cosmologies that are spatially f\/lat and homogeneous. Thus, our scale factor duality connecting homogeneous and inhomogeneous nonf\/lat cosmologies look more general. On the other hand, D\c{a}browski~et~al.\ discuss the symmetry $\gamma \rightarrow -\gamma$ of a~nonlinear oscillator equation. In our paper, the $\tilde{\gamma}$ parameter occurs mostly trigonometrically.
For the unshifted bosonic and fermionic cosmologies, we see readily that $\tilde{\gamma} \rightarrow -\tilde{\gamma}$ is a symmetry preserving their curvature indices, although we have now $u_1v_1=-\tilde{\gamma}$ and $a_g=(-\tilde{\gamma})^{-1/2\tilde{\gamma}}$, a~dif\/ferent constant.
In the case of the shifted cosmologies, the sign changes $\tilde{\gamma} \rightarrow -\tilde{\gamma}$, $\mathcal{K}_{u}(\eta;S) \rightarrow \mathcal{K}_{u}(\eta;-S)$, $\mathcal{K}_{v}(\eta;S) \rightarrow \mathcal{K}_{v}(\eta;-S)$ leave the equations unchanged.

\section{Conclusion}

In this paper, we have introduced two classes of supersymmetric-partner classes of barotropic cosmologies of Stephani type, i.e., of variable spatial curvature index, by considering a constant shift of the conformal Hubble parameter of FRW barotropic cosmologies. This has been worked out in a supersymmetric context similar to supersymmetric quantum mechanics and can be applied to any type of cosmological f\/luid unless $\gamma=\frac{2}{3}$. In other words, these new classes of inhomogeneous cosmologies together with the unshifted fermionic cosmology can be associated to any barotropic cosmological f\/luid with the exception of the coasting (non-accelerating and non-decelerating) universe.
Interestingly, we have found that an acceptable shift is purely imaginary and moreover serves as an initial condition for the conformal Hubble parameter. (It is known that `unnatural' initial conditions are required to explain why a thermodynamic arrow of time exists.) Cyclic behavior in conformal time of the cosmological zero modes can be obtained in addition to that of their curvature indices that in the pure imaginary case are also related through the parity-time (PT) property.

On the other hand, if we forget about the curvature-changing universes, the same results can be interpreted as due to the time dependence of the adiabatic indices of the cosmological f\/luids~\cite{R00}. Along this line, we notice that D\c{a}browski and Denkievicz~\cite{dd} have provided a~recent discussion
of an explicit barotropic model in which the cosmological singularity occurs only in the singular time-dependent barotropic
index. They assert that physical examples of such singularities appear in $f(R)$, scalar f\/ield, and brane cosmologies, see \cite{NO10} for a recent review in relation to all sorts of singularities. We think that the barotropic inhomogeneous cosmologies with periodic singularities in the curvature index that we introduced here could be related to the same physical examples. Moreover, the periodic singularities are not an impediment to build appropriate cosmological zero modes along the whole conformal time axis that can be used to def\/ine novel classes of scale factors corresponding to these generalized barotropic universes within any chosen period of the curvature index.

\appendix

\section[Further remarks on the solution $\mathcal{U}_{1}(\eta)$]{Further remarks on the solution $\boldsymbol{\mathcal{U}_{1}(\eta)}$}

Formula (15.3.4) in \cite{as} (in Linear transformation formulas, p.~559) reads:
\begin{gather}\label{as}
F(a,b;c;z)=(1-z)^{-a}F\left(a,c-b;c;\frac{z}{z-1}\right).
\end{gather}

Using equation~(\ref{as}) for the hypergeometric series of $\mathcal{U}_{1}(\eta)$ we get:
\begin{gather*}
{}_{2}F_{1}\left( 1,-iS
;2-iS;-e^{-2i\tilde{\gamma} \eta}\right)=
\frac{e^{i\tilde{\gamma}\eta}}{2\cos\tilde{\gamma}\eta}\,{}_{2}F_{1}\left(1,2;2-iS;\frac{1}{1+e^{2i\tilde{\gamma}\eta}}\right).
\end{gather*}
Thus:
\begin{gather}\label{pbb-1}
\mathcal{U}_{1}(\eta)=\frac{1}{2e^{S\eta}\cos \tilde{\gamma}\eta} \,{}_{2}F_{1}\left(1,2;2-iS;\frac{1}{1+e^{2i\tilde{\gamma}\eta}}\right).
\end{gather}
On the other hand, we can use the following relationship:
\begin{gather*}
(1-z)^{-p}={}_{2}F_{1}\left(p,b;b;z\right)
\end{gather*}
for $p=1$ and $S=0$ in (\ref{pbb-1}), which leads to:
\begin{gather*}
\mathcal{U}_{1}(\eta)=\frac{1}{2\cos\tilde{\gamma}\eta} 2\cos \tilde{\gamma}\eta  e^{-i\tilde{\gamma}\eta} =e^{-i\tilde{\gamma}\eta},
\end{gather*}
as expected.

\subsection*{Acknowledgements}

One of the authors (HCR) thanks the Organizing Committee for the kind invitation
and the administration of the Benasque Center for their warm hospitality. We also thank the referees for their valuable comments.

\pdfbookmark[1]{References}{ref}
\LastPageEnding

\end{document}